\documentclass[conference]{IEEEtran}
\IEEEoverridecommandlockouts
\usepackage{cite}
\usepackage[hyphens]{url} 
\usepackage{amsmath,amssymb,amsfonts}
\usepackage{graphicx}
\usepackage{textcomp}
\usepackage{xcolor}
\usepackage{algorithm}
\usepackage{algorithmicx}
\usepackage{algpseudocode}
\usepackage{graphicx}
\usepackage{booktabs}
\usepackage{subcaption}
\usepackage{caption}
\usepackage{multirow}

\def\BibTeX{{\rm B\kern-.05em{\sc i\kern-.025em b}\kern-.08em
    T\kern-.1667em\lower.7ex\hbox{E}\kern-.125emX}}
\newcommand{\oursystem}{\textit{Q\textsuperscript{2}O}}

\begin{document}

\title{Is Quantum Computing Ready for Real-Time Database Optimization?}

\author{\IEEEauthorblockN{Hanwen Liu}
\IEEEauthorblockA{\textit{Department of Computer Science} \\
\textit{University of Southern California}\\
Los Angeles, CA, USA \\
hanwen\_liu@usc.edu}
\and
\IEEEauthorblockN{Ibrahim Sabek}
\IEEEauthorblockA{\textit{Department of Computer Science} \\
\textit{University of Southern California}\\
Los Angeles, CA, USA \\
sabek@usc.edu}
} 

\maketitle

\section{Background/Question/Methods.}

\begin{figure}[h]
    \centering
    \includegraphics[width=0.8\linewidth]{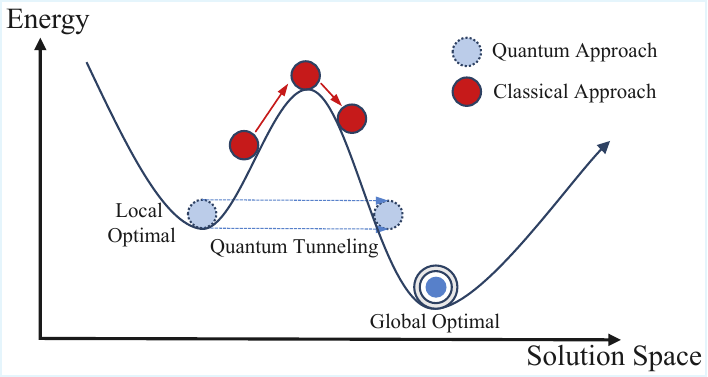}
    \caption{Quantum Annealing Compare with Classical Approach.}
    \label{fig:qa-compare}
\end{figure}

Database systems encompass several performance-critical optimization tasks, such as join ordering (e.g.,~\cite{qosurvey96,joref25,bao,yu2020reinforcementlstm,rejoin,LGS25,liu2025serag,liu2025sefrqo}) and index tuning (e.g.,~\cite{indextune1,indextune2,learnindextune22}). As data volumes grow and workloads become more complex, these problems have become exponentially harder to solve efficiently~\cite{dbtune02}. Quantum computing, especially quantum annealing~\cite{kadowaki1998}, is a promising paradigm that can efficiently explore very large search spaces~\cite{10821238,10247202} through quantum tunneling~\cite{razavy2013quantum}. As illustrated in Fig.~\ref{fig:qa-compare}, it can escape local optima by tunneling through energy barriers rather than climbing over them. Earlier works~\cite{cqmjo24,hanwenabstract,q2odemo25,transactionqubo, qtranssched25, TK16, fankhauser2021multiple, DA} mainly focused on providing an abstract representation (e.g., Quadratic Unconstrained Binary Optimization (QUBO)~\cite{Lucas2014}) for the database optimization problems (e.g., join order~\cite{qosurvey96}) and overlooked the real integration within database systems due to the high overhead of quantum computing services (e.g., a minimum 5\,s runtime for D\mbox{-}Wave’s CQM\mbox{-}Solver~\cite{dwavehss}). Recently, quantum annealing providers offered more low-latency solutions, e.g., NL\mbox{-}Solver~\cite{nlsolver25}, which paves the road to actually realize quantum solutions within DBMSs~\cite{sabekbushyjoin25}. However, this raises new systems research challenges in balancing efficiency and solution quality.

\begin{figure}[t]
    \centering
    \includegraphics[width=0.9\linewidth]{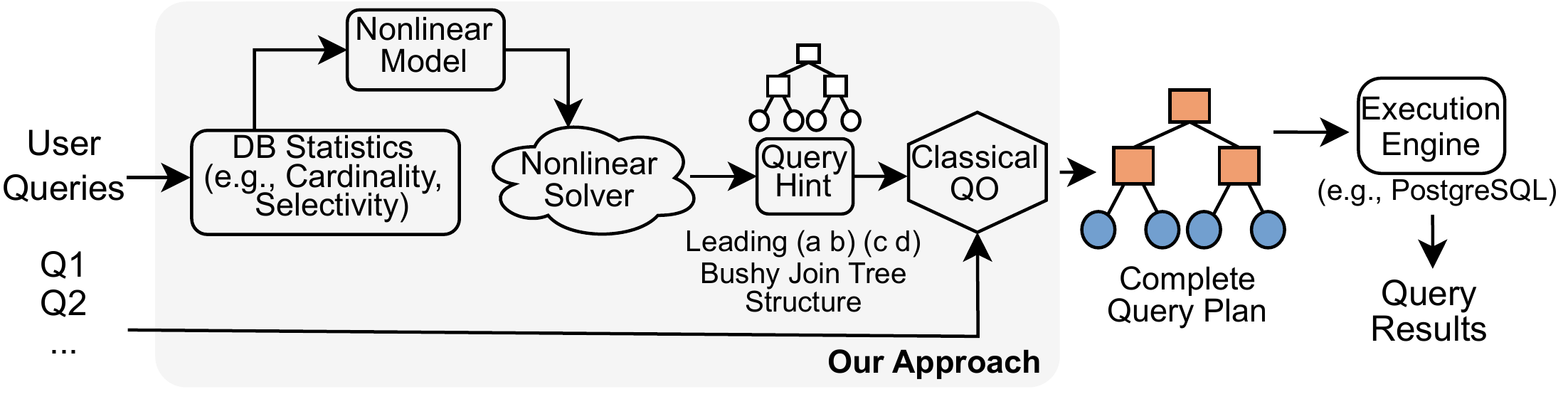}
    \caption{The Workflow of our proposed {\oursystem} Framework.}
    \label{fig:quantumdb-framework}
\end{figure}

In this talk, we show that this balance is possible to achieve. As a proof of concept, we present {\oursystem}, the \underline{\textit{first}} real {\underline{Q}uantum-augmented \underline{Q}uery \underline{O}ptimizer}. Fig.~\ref{fig:quantumdb-framework} shows the end-to-end workflow: we encode the join order problem as a nonlinear model, a format solvable by the NL\mbox{-}Solver, using actual database statistics; the solution is translated into a plan hint that guides PostgreSQL’s optimizer to produce a complete plan. {\oursystem} is capable of handling actual queries in real time.

\section{Results/Conclusions}
We set up PostgreSQL {$16.4$} with the corresponding \textit{pg\_hint\_plan} version. We evaluate {\oursystem} using the \textit{Join Order Benchmark (JOB)}~\cite{qoeval15} as real-word optimization challenges.

\begin{figure}[h]
    \centering
    \includegraphics[width=1\linewidth]{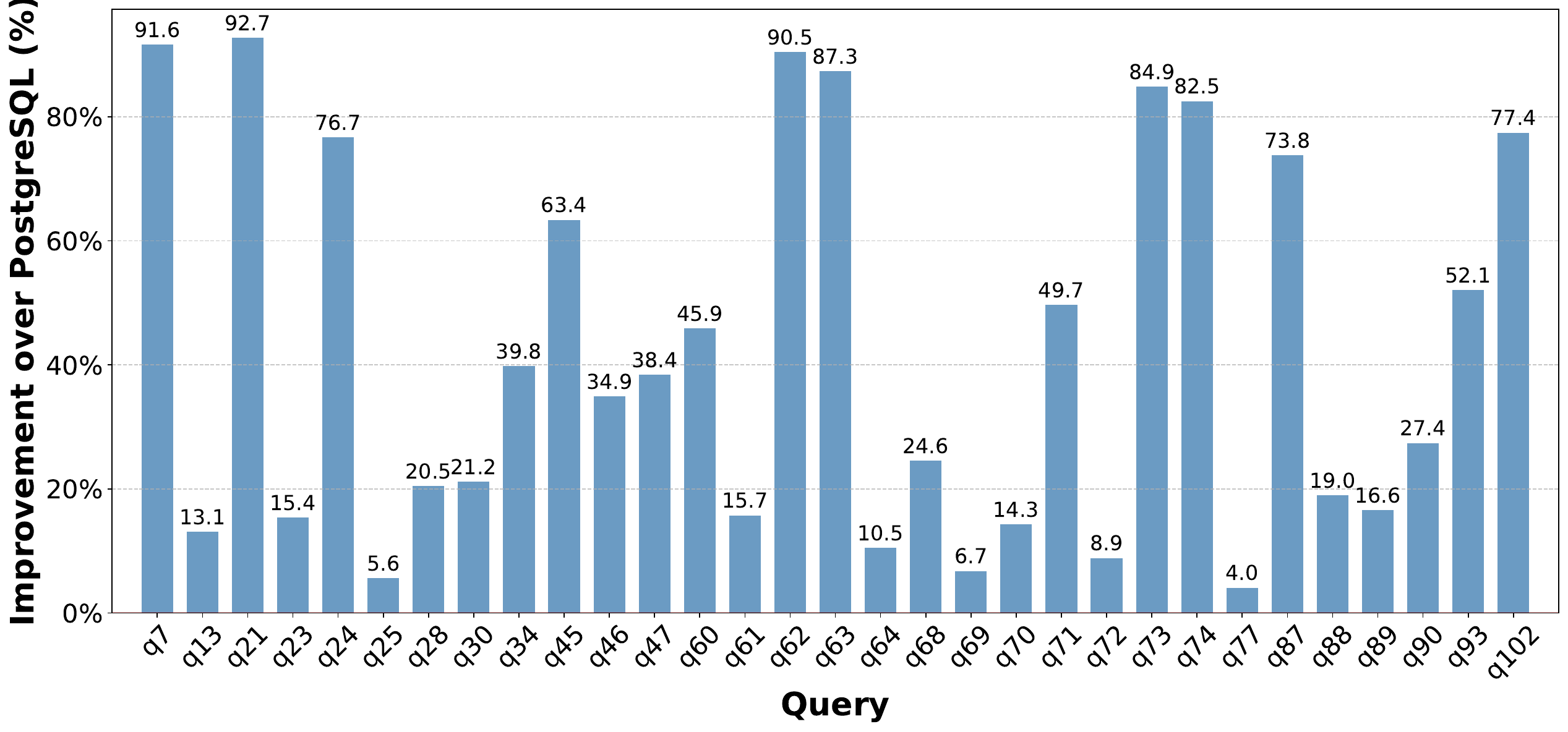}
    \caption{A Selection of Query Plans Generated by Our Approach Outperforms PostgreSQL's Default Plans.}
    \label{fig:query-plan}
\end{figure}

\noindent\textbf{Query Plan Quality.}
We evaluate the quality of the generated query plan by measuring the execution time for the JOB workload on PostgreSQL. Fig.~\ref{fig:query-plan} compares our plans with PostgreSQL’s default query plans. Among 113 queries, our method yields speedups on 31 queries, with a maximum latency reduction of $92.7\%$ and an average reduction of $42.09\%$ on those queries.

\begin{table}[t]
\setlength{\tabcolsep}{2.9pt}
\renewcommand{\arraystretch}{1.2}
\caption{End-To-End Latency Breakdown and Comparison Between Our Approach and PostgreSQL.}
\label{tab:latency}
\begin{tabular}{c|c|cccc}
\hline
 & \textbf{Time (ms)} & \textbf{q21} & \textbf{q60} & \textbf{q62} & \textbf{q63} \\ \hline
\multirow{2}{*}{\textbf{PG}} & Planning & 1.00 & 3.14 & 3.17 & 1.16 \\ \cline{2-6} 
 & Execution & 3581.40 & 10951.31 & 4680.26 & 4846.42 \\ \hline
\multirow{3}{*}{\textbf{Ours}} & Planning with hints & 2.24 & 8.31 & 9.03 & 9.02 \\ \cline{2-6} 
 & Execution & 272.35 & 6010.00 & 429.58 & 585.60 \\ \cline{2-6} 
 & NL-Solver & 2530.22 & 2748.32 & 2854.31 & 2816.42 \\ \hline
\multirow{2}{*}{\textbf{Gain}} & \textbf{Exec Time $\uparrow$} & \textbf{13.15x} & \textbf{1.82x} & \textbf{10.89x} & \textbf{8.28x} \\ \cline{2-6} 
 & \textbf{E2E Latency $\uparrow$} & \textbf{1.28x} & \textbf{1.25x} & \textbf{1.42x} & \textbf{1.42x} \\ \hline
\end{tabular}
\end{table}

\noindent\textbf{End-to-End (E2E) Latency.}
We measure E2E latency over the full pipeline. For PostgreSQL, it comprises planning time and execution time; for {\oursystem}, it includes PostgreSQL planning (with hints), PostgreSQL execution, and NL\mbox{-}Solver time. Table~\ref{tab:latency} reports per-component latencies for four JOB queries. Although the NL\mbox{-}Solver introduces additional overhead (primarily unavoidable cloud communication), our method produces higher-quality join order solutions, yields execution-time speedups up to $13.15\times$, and reduces overall E2E latency.

In conclusion, we present {\oursystem}, which leverages quantum computing to solve database optimization in real-time settings. Our evaluation demonstrates speedups in both query execution time and end-to-end latency. Looking ahead, this framework shows potential to tackle larger-scale and complex database optimization problems in real-time scenarios.
\sloppy
\bibliographystyle{IEEEtran}
\bibliography{refs/ml4db,refs/qc4db,refs/references_prof,refs/spatialmln}

@string{VLDB="Proceedings of the International Conference on Very Large Data Bases, VLDB"}

@string{SIGMOD="Proceedings of the ACM International Conference on Management of Data, SIGMOD"}

@string{ICDE="Proceedings of the International Conference on Data Engineering, ICDE"}

@string{IDEAS="Proceedings of the International Database Engineering and Applications Symposium, IDEAS"}

@string{SIGMOD="SIGMOD"}

@string{VLDB="VLDB"}

@string{ICDE="IEEE ICDE"}

@string{IDEAS="IDEAS"}

@inproceedings{bao,
no_author = {R. Marcus and others},
author = {Marcus, Ryan and Negi, Parimarjan and Mao, Hongzi and Tatbul, Nesime and Alizadeh, Mohammad and Kraska, Tim},
title = {{Bao: Making Learned Query Optimization Practical}},
year = {2021},
booktitle = SIGMOD
}

@article{qoeval15,
author = {Leis, Viktor and Gubichev, Andrey and Mirchev, Atanas and Boncz, Peter and Kemper, Alfons and Neumann, Thomas},
title = {How Good Are Query Optimizers, Really?},
year = {2015},
issue_date = {November 2015},
publisher = {VLDB Endowment},
volume = {9},
number = {3},
issn = {2150-8097},
url = {https://doi.org/10.14778/2850583.2850594},
doi = {10.14778/2850583.2850594},
journal = {Proc. VLDB Endow.},
month = {nov},
pages = {204–215},
numpages = {12}
}

@misc{liu2025sefrqo,
      title={SEFRQO: A Self-Evolving Fine-Tuned RAG-Based Query Optimizer}, 
      author={Hanwen Liu and Qihan Zhang and Ryan Marcus and Ibrahim Sabek},
      year={2025},
      eprint={2508.17556},
      archivePrefix={arXiv},
      primaryClass={cs.DB},
      url={https://arxiv.org/abs/2508.17556}, 
}

@inproceedings{LGS25,
  author    = {Hanwen Liu and Shashank Giridhara and Ibrahim Sabek},
  title     = {{Conformal Prediction for Verifiable Learned Query Optimization}},
  booktitle = {Proceedings of the International Conference on Very Large Data Bases, VLDB},
  address	= {London, United Kingdom},
  year      = 2025
}

@misc{liu2025serag,
      title={SERAG: Self-Evolving RAG System for Query Optimization}, 
      author={Hanwen Liu and Qihan Zhang and Ryan Marcus and Ibrahim Sabek},
      year={2025},
      url={https://viterbi-web.usc.edu/~sabek/pdf/25_workshop_serag.pdf}, 
}

@inproceedings{learnindextune22,
author = {Wu, Wentao and Wang, Chi and Siddiqui, Tarique and Wang, Junxiong and Narasayya, Vivek and Chaudhuri, Surajit and Bernstein, Philip A.},
title = {Budget-aware Index Tuning with Reinforcement Learning},
year = {2022},
isbn = {9781450392495},
publisher = {Association for Computing Machinery},
address = {New York, NY, USA},
url = {https://doi.org/10.1145/3514221.3526128},
doi = {10.1145/3514221.3526128},
booktitle = {Proceedings of the 2022 International Conference on Management of Data},
pages = {1528–1541},
numpages = {14},
location = {Philadelphia, PA, USA},
series = {SIGMOD '22}
}

@inproceedings{yu2020reinforcementlstm,
  title={Reinforcement learning with tree-lstm for join order selection},
  author={Yu, Xiang and Li, Guoliang and Chai, Chengliang and Tang, Nan},
  booktitle={2020 IEEE 36th international conference on data engineering (ICDE)},
  pages={1297--1308},
  year={2020},
  organization={IEEE}
}

@inproceedings{TK16,
author = {Immanuel Trummer and Christoph Koch},
title = {{Multiple Query Optimization on the D-Wave 2X Adiabatic Quantum Computer}},
year = {2016},
no_issue_date = {May 2016},
no_publisher = {VLDB Endowment},
no_volume = {9},
no_number = {9},
no_issn = {2150-8097},
no_url = {https://doi.org/10.14778/2947618.2947621},
no_doi = {10.14778/2947618.2947621},
booktitle = VLDB,
no_month = {may},
no_pages = {648–659},
no_numpages = {12}
}

@inproceedings{DA,
author = {Manuel Sch\"{o}nberger and Immanuel Trummer and Wolfgang Mauerer},
title = {{Quantum-Inspired Digital Annealing for Join Ordering}},
year = {2023},
no_issue_date = {November 2023},
no_publisher = {VLDB Endowment},
no_volume = {17},
no_number = {3},
no_issn = {2150-8097},
no_url = {https://doi.org/10.14778/3632093.3632112},
no_doi = {10.14778/3632093.3632112},
booktitle = VLDB,
no_month = {nov},
no_pages = {511–524},
no_numpages = {14}
}

@misc{dwavehss,
  author =        {{Hybrid Solver for Constrained Quadratic Models [WhitePaper]}},
  no_key =        {{Hybrid Solver for Constrained Quadratic Models [WhitePaper]}},  
  no_title =        {{Hybrid Solver for Constrained Quadratic Models [WhitePaper]}},
  year = 2021,
  howpublished =          "\url{https://www.dwavesys.com/media/rldh2ghw/14-1055a-a_hybrid_solver_for_constrained_quadratic_models.pdf}"
}

@article{lucas2014,
	author = {Lucas, Andrew},
	doi = {10.3389/fphy.2014.00005},
	issn = {2296-424X},
	journal = {Frontiers in Physics},
	title = {Ising formulations of many NP problems},
	url = {https://www.frontiersin.org/articles/10.3389/fphy.2014.00005},
	volume = {2},
	year = {2014}}

@inproceedings{rejoin,
author = {Marcus, Ryan and Papaemmanouil, Olga},
title = {Deep Reinforcement Learning for Join Order Enumeration},
year = {2018},
isbn = {9781450358514},
publisher = {Association for Computing Machinery},
address = {New York, NY, USA},
url = {https://doi.org/10.1145/3211954.3211957},
doi = {10.1145/3211954.3211957},
booktitle = {Proceedings of the First International Workshop on Exploiting Artificial Intelligence Techniques for Data Management},
articleno = {3},
numpages = {4},
location = {Houston, TX, USA},
series = {aiDM'18}
}

@inproceedings{transactionqubo,
author = {Groppe, Sven and Groppe, Jinghua},
title = {Optimizing Transaction Schedules on Universal Quantum Computers via Code Generation for Grover’s Search Algorithm},
year = {2021},
isbn = {9781450389914},
publisher = {Association for Computing Machinery},
address = {New York, NY, USA},
url = {https://doi.org/10.1145/3472163.3472164},
doi = {10.1145/3472163.3472164},
booktitle = {Proceedings of the 25th International Database Engineering \& Applications Symposium},
pages = {149–156},
numpages = {8},
location = {Montreal, QC, Canada},
series = {IDEAS '21}
}

@article{kadowaki1998,
  title = {Quantum annealing in the transverse Ising model},
  author = {Kadowaki, Tadashi and Nishimori, Hidetoshi},
  journal = {Phys. Rev. E},
  volume = {58},
  issue = {5},
  pages = {5355--5363},
  numpages = {0},
  year = {1998},
  month = {Nov},
  publisher = {American Physical Society},
  doi = {10.1103/PhysRevE.58.5355},
  url = {https://link.aps.org/doi/10.1103/PhysRevE.58.5355}
}

@article{fankhauser2021multiple,
  title={Multiple query optimization using a hybrid approach of classical and quantum computing},
  author={Fankhauser, Tobias and Sol{\`e}r, Marc E and F{\"u}chslin, Rudolf M and Stockinger, Kurt},
  journal={arXiv preprint arXiv:2107.10508},
  year={2021}
}

@inproceedings{hanwenabstract,
author = {Liu, Hanwen and Saxena, Pranshi and Spedalieri, Federico and Sabek, Ibrahim},
title = {Optimizing Join Orders via Constrained Quadratic Models (Abstract)},
year = {2025},
booktitle = {Proc. 1st Workshop Quantum Data and Machine Learning},
location = {HongKong, China},
series = {QDML '25}
}

@book{razavy2013quantum,
  title={Quantum theory of tunneling},
  author={Razavy, Mohsen},
  year={2013},
  publisher={World Scientific}
}

@string{ICDE="Proceedings of the IEEE International Conference on Data Engineering, IEEE ICDE"}

@string{CHI="Processing of the ACM International Conference on Human Factors in Computing Systems, CHI"}

@string{ICDE="ICDE"}

\end{document}